\documentclass[structabstract]{aa}  
\usepackage{natbib}
\usepackage{graphicx}
\usepackage{color}
\usepackage{txfonts}
\definecolor{lila}{rgb}{0.5,0,1}
\definecolor{grau}{rgb}{0.5,0.5,0.5}

\newcommand{\Tpeff}{T_{p,\,\rm eff}}

\newcommand{\Tstar}{T_{\ast,\,\rm eff}}
\newcommand{\Mstar}{M_{\ast}}
\newcommand{\Rstar}{R_{\ast}}
\newcommand{\ME}{M_{\oplus}}
\newcommand{\RE}{R_{\oplus}}

\begin{document}
\title{Interior structure models of GJ$\:$436b}

\author{N. Nettelmann\inst{1,2}\fnmsep
        U. Kramm\inst{1}, R. Redmer\inst{1},\and 
        R. Neuh\"auser\inst{3}\fnmsep
}
\institute{Institut f\"ur Physik, Universit\"at Rostock, Universit\"atsplatz 3, 18051 Rostock, Germany
\and Department of Astronomy and Astrophysics, University of California, Santa Cruz, CA 95064, U.S.A. 
\and Astrophysikalisches Institut und Universit\"ats-Sternwarte Friedrich-Schiller-Universit\"at 
Jena, Schillerg\"asschen 2-3, 07745 Jena, Germany
}
\date{Received 4 March 2009, accepted 6 August 2010}

\abstract
{GJ$\:$436b is the first extrasolar planet discovered that resembles Neptune in mass and radius. 
Two more are known (HAT-P-11b and Kepler-4b), and many more are expected to be found in 
the upcoming years. The particularly interesting property of Neptune-sized planets is that their 
mass $M_p$ and radius $R_p$ are close to theoretical $M$-$R$ relations of water planets. 
Given $M_p$, $R_p$, and equilibrium temperature, however, various internal compositions are possible.}
{A broad set of interior structure models is presented here that illustrates the dependence
of internal composition and possible phases of water occurring in presumably water-rich planets, 
such  as GJ$\:$436b on the uncertainty in atmospheric temperature profile and mean density. 
We show how the set of solutions can be narrowed down if theoretical constraints from formation 
and model atmospheres are applied or potentially observational constraints 
for the atmospheric metallicity $Z_1$ and the tidal Love number $k_2$.}
{We model the interior by assuming either three layers (hydrogen-helium envelope, water layer, rock core)
or two layers (H/He/H$_2$O envelope, rocky core). For water, we use the equation of state 
H$_2$O-REOS based on finite temperature - density functional theory - molecular dynamics (FT-DFT-MD) simulations.}
{Some admixture of H/He appears mandatory for explaining the measured radius. For the warmest 
considered models, the H/He mass fraction can reduce to $10^{-3}$, still extending 
over $\sim 0.7\RE$. If water occurs, it will be essentially in the plasma phase or in the 
superionic phase, but not in an ice phase. Metal-free envelope models have $0.02<k_2<0.2$,
and the core mass cannot be determined from a measurement of $k_2$. In contrast, 
models with $0.3<k_2<0.82$ require high metallicities $Z_1<0.89$ in the outer envelope.
The uncertainty in core mass decreases to $0.4M_p$, if $k_2\geq 0.3$, and further to $0.2 M_p$, 
if $k_2\geq 0.5$, and core mass and $Z_1$ become sensitive functions of $k_2$.}
{To further narrow the set of solutions, a proper treatment of the atmosphere and the 
evolution is necessary. We encourage efforts to observationally determine the atmospheric 
metallicity and the Love number $k_2$.}

\keywords{planets and satellites: interiors -- planets and satellites: individual: GJ$\:$436b 
-- EOS: water equation of state}
\authorrunning{Nettelmann et al.}
\titlerunning{Interior models of GJ$\:$436b}
\maketitle

\section{Introduction}

The planet GJ$\:$436b~\citep{Butler+04} has attracted much attention 
(e.g.~\citealp{Maness+07,Baraffe+08,Batygin+09a}) because it is the first 
Neptune-mass extrasolar planet observed transiting its parent 
star~\citep{Gillon+07b}. Follow-up Doppler observations and infrared photometry data 
from primary transit and secondary eclipse~\citep{Demory+07,Deming+07}
have allowed its mass $M_p$, radius $R_p$, and day-side brightness 
temperature to be determined to within 14\%, 13\% and 25\%, respectively. 
For Neptune-size extrasolar planets, these parameters permit a variety of 
possible compositions ranging from a rock-iron planet surrounded by a gaseous 
H/He layer to a planet that is substantially composed of water~\citep{Adams+08,Fortney+07a}. 
For instance, for the Neptune-size solar planets Uranus and Neptune, 
more observational constraints such as the gravitational moments are available, 
but their bulk composition is still unknown due to the ambiguity of the mean density of 
water and H/He/rock mixtures~\citep{PPM-UN00}.

\citet{RogSea10a} quantify the uncertainty in the composition of Neptune-mass planets by 
taking the uncertainties in $M_p$, $R_p$, and intrinsic luminosity into account. For 
GJ$\:$436b, they obtain mass fractions of H/He, water, and rocks, where rocks comprise 
silicates and iron, of 3.6$-$14.5\%, 0$-$96.4\%, and 4$-$90\%, respectively. They find a 
particularly strong dependence of composition on the atmospheric thermal profile. 
\citet{Figueira+09} are able to further constrain the composition by comparing with possible 
formation histories. Their formation models exclude water-less models and predict initial 
mass fractions of H/He, ice, and rock of 10-20\%, 45-70\%, and 17-40\%, respectively.
This demonstrates the necessity of further constraints to distinguish between 
various interior solutions. Such an additional observable would be the tidal Love number 
$k_2$~\citep{RW09,Batygin+09b}. 
For the two-planet system HAT-P13b,c~,~\citet{Batygin+09b} show how high-precision 
measurements of the orbital and planetary parameters can translate into information about 
the planetary interior of the inner planet, such as parametrized in terms of a core mass.
GJ$\:$436b's high eccentricity suggests the presence of such an additional companion 
\citep{Ribas+08}. Predictions for the perturbers's mass of $\approx 10\ME$ 
\citep{Coughlin+08,Batygin+09a} and eccentricity are just within the scope of current 
detection thresholds \citep{Batygin+09a}, so the value of the observable $k_2$ 
could be inferred. In this paper we discuss the information content of $k_2$ with respect to 
the core mass and metallicity of GJ$\:$436b, and its implication for a secondary planet. 

Core mass and metallicity are calculated for a broad range of models that aim to embrace the 
full set of solutions that are allowed by the uncertainty in $M_p$, $R_p$ and the atmospheric 
temperature profile. The material components that GJ$\:$436b is assumed to be made of are rocks, 
confined to a rocky core, water, whether confined to an inner envelope or uniformly mixed 
into an outer hydrogen-helium envelope, and hydrogen and helium. 
For H, He, and water, we apply the Linear Mixing Rostock Equation Of State (LM-REOS) described in 
\citet{N-Jupiter}, which is based on finite temperature - density functional theory - 
molecular dynamics (FT-DFT-MD) simulations for the components H, He, and H$_2$O, 
\citep[see][]{Holst+08,Kietz+07,French+09}. In particular, the phase diagram of water has been 
calculated recently up to pressures of 100~Mbar and temperatures of 40000~K \citep{French+09}, 
so that we can derive the possible phases of water in presumably water-rich planets 
such as GJ$\:$436b in dependence on the uncertainties mentioned above. 

We describe and discuss the observational constraints applied in \S$\,$\ref{ssec:meth_obs},
the equation of state data in \S$\,$\ref{ssec:meth_EOS}, and our method of generating 
structure models in \S$\,$\ref{ssec:meth_structure}. 
The range of resulting compositions are presented and discussed in \S$\,$\ref{ssec:res_waterphases} 
in the context of the phase diagram of water and constraints from formation theory and  
model atmospheres. The range of resulting core masses and metallicities is presented and 
discussed in \S$\,$\ref{ssec:res_k2} together with the potentially observable Love number $k_2$.
We find that, at deep interior temperatures and pressures in GJ$\:$436b, 
finite-temperature effects of the water EOS are important. In \S$\,$\ref{ssec:MRrels} 
we present new $M$-$R$ relations of warm water planets. Our conclusions are discussed and 
summarized in \S$\,$\ref{sec:discussion}.

\section{Input data and model construction}\label{sec:methods}

\subsection{Observational constraints}\label{ssec:meth_obs}

Observational data of the mass $M_p$, radius $R_p$, and surface temperature of GJ$\:$436b 
have been provided by serveral authors as summarized below.
It is interesting to note that different determinations of $M_p$ and $R_p$ arise 
mostly from different assumptions about the stellar mass $\Mstar$ and radius $\Rstar$. 

\citet{Deming+07} determined a planet-to-star radius ratio $R_p/\Rstar$ of $0.0839\pm0.0005$
from the depth in the precise \emph{Spitzer} transit photometry light curve. 
They then used the \emph{Spitzer} light curve and radial velocity data to find the best 
fit to the stellar radius versus the stellar mass and compared this relation to the empirical 
$M$-$R$ relation $\Mstar=\Rstar$ to find $\Mstar=\Rstar=0.47 \pm 0.02$ (in solar units) 
and consequently a planet radius $R_p=4.33\pm 0.18$ Earth radii ($\RE$). From the 
amplitude of the secondary eclipse observed with \emph{Spitzer}, they also determined 
a day-side brightness temperature of the planet of $712 \pm 36$~K. They also give 
$M_p=0.070 \pm 0.003$ Jupiter masses $(M_{\rm J})$ as best fit to all observations.
Assuming $\Mstar=0.44M_{\odot}$ and $\Rstar=0.463R_{\odot}$, \citet{Gillon+07a,Gillon+07b} 
and \citet{Demory+07} find similar planetary values within the error bars by analyzing 
the \emph{Spitzer} transit data. 

\citet{Torres+08} use more recent results from stellar evolution models that match the 
observed stellar luminosity derived from spectroscopic measurements of the effective 
stellar temperature and the mean density derived from photometric measurements 
of $a/\Rstar$, where $a$ is the planet-star separation. They obtain as best values 
$\Mstar=0.452\pm 0.014M_{\odot}$ and $\Rstar=0.464\pm 0.011 R_{\odot}$ for GJ$\:$436 with
error bars covering previous estimates, and  $M_p=0.0729\pm 0.0025M_{\rm J}$, and 
$R_p=0.3767^{+0.0082}_{-0.0092} R_{\rm J}$ for GJ$\:$436b, where $R_{\rm J}$ is the 
equatorial Jupiter radius. \citet{Torres+08} and also \citet{Deming+07} assume the effective 
temperature $\Tstar$ of GJ$\:$436 to be $3350\pm 300$~K, which we think is too low and not
constrained enough for an M3V star with a V-K color index of 4.7 mag (V=10.68, Simbad; 
K=6.073, 2MASS). For such an M star, we obtain $\Tstar=3470\pm 100$~K from \citet{KH95} in 
agreement with \citet{Bean+06}, who compared their optical high-resolution spectra of GJ$\:$436 
with the latest PHOENIX model atmospheres to obtain $\Tstar=3480$~K. This leads to a slightly 
higher planet equilibrium temperature $T_{eq}=\Tstar(\Rstar/2a)^{1/2}$ of $673\pm 20$~K 
instead of $649\pm 60$~K, assuming zero-albedo. Within the error bars, however, the equilibrium
temperatures agree with each other and also with the planet's effective temperature 
($\Tpeff=712\pm 36$~K, \citealt{Deming+07}; $717\pm 35$~K, \citealt{Demory+07}) obtained 
from \emph{Spitzer} infrared data, but no longer does so if assuming a non-zero albedo $a_p=0.3$ 
as of Uranus yielding $T_{eq}=535$~K, which would then imply a significant intrinsic luminosity. 

\citet{Bean+08} have repeatedly observed the transit with the Hubble Space Telescope (HST) 
Fine Guidance Sensor (FGS) and fitted the transit light curves to directly obtain 
the stellar radius, as well as the planetary radius, orbital period, and inclination. 
They find $R_p=4.90^{+0.45}_{-0.33}\RE$, i.e.~larger than all other radii measured by others. 
Our baseline models have $M_p=23.17\ME$ and $R_p=4.22\RE$ in accordance with \citet{Torres+08}.

\subsection{EOS}\label{ssec:meth_EOS}

Interior models presented in this work are composed of H, He, H$_2$O, and rocks. 
The rock-EOS used here is an analytical pressure-density relation by \citet{HM89} 
that approximates a theoretical EOS of a mixture of 38\%~$\rm SiO_2$, 25\%~$\rm MgO$, 
25\%~$\rm FeS$, and 12\%~$\rm FeO$ appropriate for Jovian core conditions, i.e.~for 
$T\sim10^4$~K. This rock-EOS has been applied extensively to the core region of the 
solar system giant planets \citep{HM89,Guillot+94}. 

Mixtures of H, He, and H$_2$O are generated by linear mixing. For these components we use 
H-, He-, and H$_2$O-REOS, respectively, which are based on FT-DFT-MD simulations at those 
densities where correlation effects are important and on chemical models in complementary 
regions, see \citet{N-Jupiter} and references therein for a more detailed description. 
In particular, H$_2$O-REOS is a combination of different finite-temperature EOS covering phases 
ice~I and liquid water by fits to accurate experimental data, water vapor using Sesame EOS 
7150~\citep{SESAME}, supercritical molecular water (Sesame 7150 and FT-DFT-MD), 
fractions of ice {\small VII and X}, and large regions of superionic water and water plasma 
up to 20~g~cm$^{-3}$ and 40000~K (FT-DFT-MD). All interior models presented here are too warm 
for ice phases to occur. While real planets may have accreted water in combination with 
methane and ammonia (together conveniently named \emph{ices} regardless of their thermodynamic 
phase at planetary interior conditions), we consider the ab initio EOS H$_2$O-REOS an improvement 
over former \emph{ice}-EOS, because temperature effects are properly taken into account.

For comparison, \citet{Adams+08} and \citet{RogSea10a} use EOS data by \citet{Seager+07} 
for their GJ$\:$436b models, where the iron and silicate EOS reproduce experimental data 
at $P\leq 2$~Mbar and the water EOS is in addition based on DFT-MD data for water-ice 
phases {\small VIII and X}. \citet{Baraffe+08} extensively studied the effect of 
different available water EOS, i.e. of Sesame and ANEOS water EOS. Isotherms of these EOS 
show a significantly more compressible behavior in the Mbar region than the new H$_2$O-REOS, 
see \citet{French+09}, thereby requiring a higher mass fraction of the light H/He 
component to yield the same planetary mean density.

\subsection{Structure assumptions}\label{ssec:meth_structure}

We obtained interior models by integrating the equation of hydrostatic equilibrium of a gravitating 
sphere $dP/dr=-Gm(r)\rho(r)/r^2$ and the equation of mass conservation $dm/dr=4\pi r^2\rho(r)$ inward, 
where $P$ is pressure, $\rho$ mass density, $m$ the mass, and $r$ the radial coordinate. 
These equations require three conditions at the outer boundary $r_1=R_p$ for the variables 
$m,P$, and $\rho$. Clearly, $m(r_1)=M_p$. In accordance with prior interior structure models of 
solar giant planets (e.g. \citealt{Steve82}), we define the surface as the 1-bar pressure level, 
thus $P(r_1)=1$~bar. For $\rho_1=\rho(P_1,T_1)$ from the EOS we need the thermodynamical  surface 
temperature $T_1$ which we consider a variable parameter. Since a real planet has to obey the 
inner boundary condition $m(0)=0$, we make the additional assumption that a core exists and choose 
the respective core mass $M_{core}$ such that total mass conservation is ensured. We define a layer 
by its homogeneous composition, where  pressure $P_{n-(n+1)}$ and temperature at the transition 
from layer No.~$n$ to No.~$n+1$ are continuous, while density and entropy change discontinuously. 

In this work we consider two-layer (2L) and three-layer (3L) models. 
For our 3L models, the innermost layer No.~3 is a core of rocks, layer No.~2 is an 
adiabatic inner envelope of water with envelope metallicity $Z_2=1$, and layer No.~1 is an 
outer envelope of linearly mixed hydrogen and helium with an He mass fraction 
$Y=M_{\rm He}/(M_{\rm H}+M_{\rm He})=0.27$ as is characteristic for the protosolar cloud and 
envelope metallicity $Z_1=0$. As in \citet{Adams+08}, we assume an adiabatic profile for 
$P>1$~kbar that is located in the outer envelope. This structure type is in line with 
pioneering Uranus and Neptune models by \citet{HMacF80}, who investigated a layered interior 
differentiated into a central rocky (silicates+iron) core overlaid by an ice shell 
(H$_2$O, CH$_4$, NH$_3$) and surrounded by a H/He envelope. 
While such models do not reproduce the gravitational moments $J_2$ and $J_4$ of Uranus and 
Neptune \citep{Hubb+95}, implying an admixture of heavy elements into the outer H/He envelope, 
as well as some admixture of H/He into the inner, ice-rich envelope, they are a reasonable 
starting point when additional gravity field data are not at hand. 
Given only $M_p$, $R_p$, and $\Tpeff$ and a location of GJ$\:$436b close to theoretical 
$M$-$R$ relations for pure water planets, we expect a broad range of solutions even if 
observational uncertainties in $M_p$ and $R_p$ are neglegted. For our 3L structure type 
models, we vary the transition pressure $P_{1-2}$ from the outer to the inner envelope 
between the bounds as defined by zero-mass water layer and zero-mass rocky core.

Alternatively, we calculated two-layer models where water is homogeneously mixed into 
the H/He envelope. For this structure type, we varied the envelope metallicity 
$Z_1:=M_{\rm H_2O, env}/M_{\rm env}$ between the bounds as defined by $Z_1=0$ and a zero-mass
core on the other hand. These two kinds of simple structure assumptions bracket the uncertainty 
regarding the distribution of ices within the envelopes. The presence of heavier elements such 
as silicates in the H/He/H$_2$O envelope or in the water layer of the 3L models would act to 
reduce the amount of water in favor of the amount of H/He needed to reproduce the radius. 
These models therefore give a lower limit for the H/He mass fraction in GJ$\:$436b.

An important question is the temperature profile between 1 and 1000~bar, 
where atmospheric models for irradiated planets based on non-gray radiative-convective 
equilibrium calculations \citep{Fortney+07a,Burrows+08} predict an isothermal 
region. Adiabatic atmospheres, on the other hand, are generally assumed for all solar giant 
planets owing to weak irradiation and strong molecular absorption in that pressure region 
\citep{Guillot+94} making energy transport by radiation inefficient. 
For Jupiter, this assumption is confirmed by Galileo probe entry measurements and for 
Uranus \& Neptune by \emph{Voyager} infrared and radio occultation observations 
\citep{Gautier+95}. Since we aim to not narrow the possible set of solutions in advance 
by the choice of a specific model atmosphere, we cover these limits in the case 
of our 3L models by both calculating isothermal and adiabatic models in the pressure 
range 1 to 1000~bar representing the extreme cases of strong or no influence of 
irradiation. For both 2L and 3L models, we vary $T_1$ between 300~K and 1300~K. 
Cold interior models of GJ$\:$436b, in particular with water cores at 300~K overlaid by 
an adiabatic, hence an even colder, H/He atmosphere have been studied by \citet{Baraffe+08} 
with respect to the evolution timescale. 
\citet{Spiegel+10}, on the other hand, find isothermal atmospheres at 1300~K and $1-100$~bar, 
independent of the day-to nightside energy redistribution in the atmosphere of GJ$\:$436b.
For our 2L models, we also vary $T_1:=T$(1~bar) from 300 to 1300~K but only consider 
temperature profiles with $T_1$ = $T_{100}$ constant, where $T_{100}:=T$(100~bar), and 
adiabatic for $P>100$~bar. 

We subdivide our structure type models into 11 series labeled A, B,\ldots, K representing 
the uncertainties in $T_1$, $T_{1000}$, and mean density $\bar{\rho}_p$, where 
$T_{1000}:=T$(1000 bar). For series A$-$I, we take $M_p=0.0729M_{\rm J}=23.17\ME$ and 
$R_p=0.3767R_{\rm J}=4.22\RE$; for series J, we take $M_p=21.3\ME$ and $R_p=5.35\RE$ 
corresponding to the minimum value of $\bar{\rho}_p$; for series K, we take $M_p=24\ME$ 
and $R_p=4\RE$ representing a maximal $\bar{\rho}_p$. 
They are modifications of series E intended to investigate the effect of the mean density. 
The input characteristics of our model series A$-$K are listed in Table~\ref{tab:series}.

\begin{table}[hbt]
\caption{\label{tab:series}Input characteristics of model series A$-$K.}
\centering
\begin{minipage}{\hsize}
\centering{
\begin{tabular}{cccccc}
series	& $T_1$		& $\nabla_T$						& $M_p$	 	& $R_p$		& N \\
			& [K]		&											& [$\ME$]	&[$\RE$]	& (layers)\\
\hline\hline		
A			& 300		& adiabatic							& 23.17		& 4.22		& 3\\
B			& 300		& isothermal 1-10$^3$~bar	& 23.17		& 4.22		& 3\hfill\\
C			& 300		& isothermal 1-10$^2$~bar	& 23.17		& 4.22		& 2\\
D			& 700		& adiabatic							& 23.17		& 4.22		& 3\\
E			& 700		& isothermal 1-10$^3$~bar	& 23.17		& 4.22		& 3\hfill\\
F			& 700		& isothermal 1-10$^2$~bar	& 23.17		& 4.22		& 2\\
G			& 1300		& adiabatic							& 23.17		& 4.22		& 3\\
H			& 1300		& isothermal 1-10$^3$~bar	& 23.17		& 4.22		& 3\hfill\\
I			& 1300		& isothermal 1-10$^2$~bar	& 23.17		& 4.22		& 2\\
J			& 700		& isothermal 1-10$^3$~bar	& 21.3		& 5.35		& 3\\
K			& 700		& isothermal 1-10$^3$~bar	& 24.0		& 4.0		& 3\\\hline
\end{tabular}}
\end{minipage}
For series A$-$I, $M_p$ and $R_p$ are taken from \citet{Torres+08}, yielding a mean density 
$\bar{\rho}_p=1.69\rm~g~cm^{-3}$, while in series J, $\bar{\rho}_p=0.45\rm~g~cm^{-3}$, and in series K, 
$\bar{\rho}_p=2.22\rm~g~cm^{-3}$.
\end{table}

\subsubsection{Love number $k_2$}

For selected models, we calculated the Love number $k_2$. This planetary property quantifies 
the strength of the planet's quadrupolic gravity field deformation in response to an external 
massive body. In case of the system GJ$\:$436 and GJ$\:$436b, $\Mstar$ causes a tide-raising 
potential $W(r)=\sum_{n=2}W_n=(GM_*/a)\sum_{n=2}(r/a)^nP_n(\cos\psi)$, where $a$ is the 
(time-dependent) distance of the center of masses, $r$ the radial coordinate of a 
point inside the planet in a planet-centered coordinate system, $\psi$ the angle between 
a planetary mass element at $\vec{r}$ and $M_*$ at $\vec{a}$, and $P_n$ are Legendre 
polynomials. Each external potential's pole moment $W_n$ induces a disturbance $V_n=K_n(r) W_n$ 
of the planetary potential, with $k_n=K_n(R_p)$ taken at the surface $R_p$ the Love numbers. 
Relevant equations to calculate Love numbers are given in \citet[\S 48]{ZT78}. 

As the gravitational moments $J_{2n}$, the Love numbers are uniquely defined by the 
planet's internal density distribution. To first order in the expansion of a planet's 
potential, $k_2$ is proportional to $J_2$ \citep[see e.g.][]{HubbBuch}, so that  
measuring $k_2$ provides us with an additional constraint equivalent to $J_2$ for 
the solar system giant planets. In particular,  $k_2$ is --as is $J_2$-- most sensitive 
to the central condensation of a planet, as parametrized adequately by the mass of the core 
within a two-layer approach. However, as $J_2$ and $k_2$ depend on the accumulated density
distribution $\rho(r)$, the inverse process of deriving $\rho(r)$ from $J_2$ or $k_2$ is 
not unique. Even in the most fortunate case of enough moments being available to accurately 
determine the internal density distribution --which is still uncertain even for Jupiter 
\citep{ForNet09}--  we have to face the ambiguity from various materials yielding the 
same density when mixed. For example, the common perception of Uranus and Neptune as 
mostly made of ices is supported by the observed atmospheric strong enrichment in methane 
\citep{Hubb+95}, but ice-free interior models are also allowed by the gravity field data
\citep{PPM-UN00}. We present results for the information content of $k_2$ based on our 
model series A$-$I in~\S~\ref{ssec:res_k2}.

\section{Results}\label{sec:results}

\subsection{Water in the interior of GJ$\:$436b}\label{ssec:res_waterphases}

\begin{figure*}[hbt]
\flushleft
\includegraphics[angle=0,width=4cm]{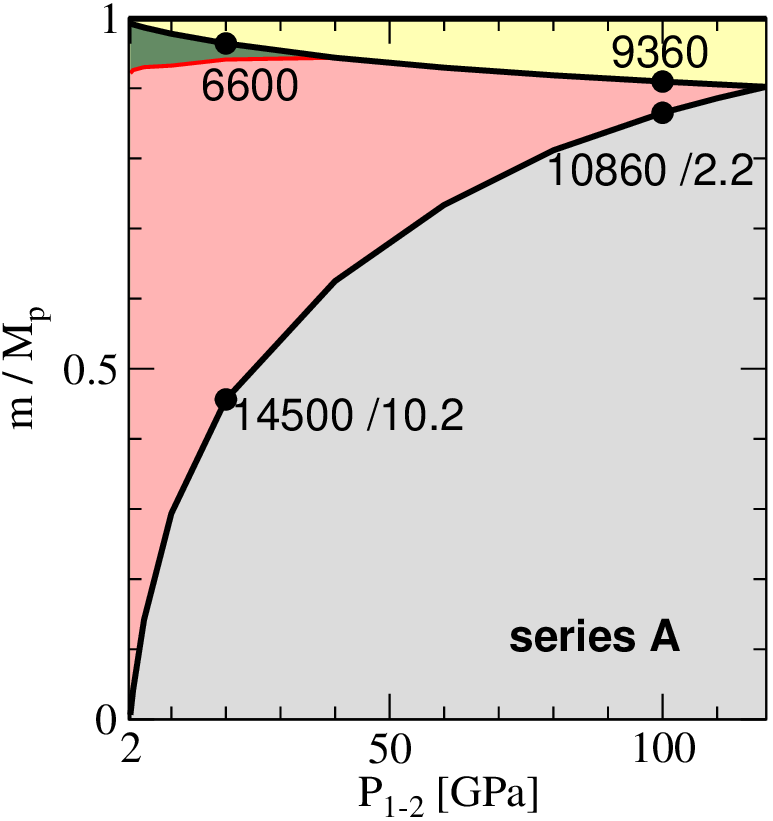}
\includegraphics[angle=0,width=4cm]{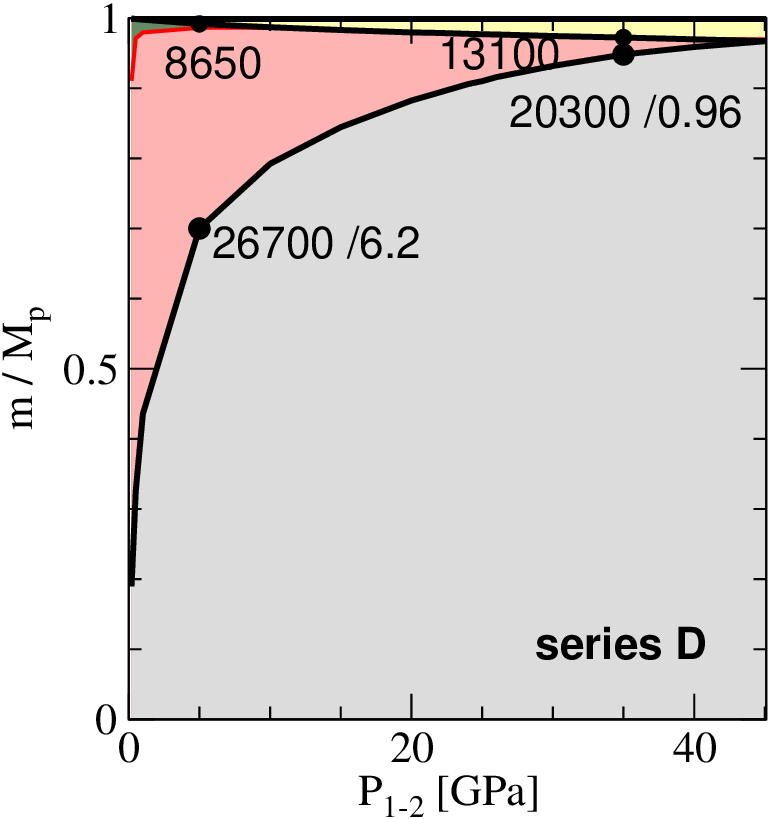}
\includegraphics[angle=0,width=4cm]{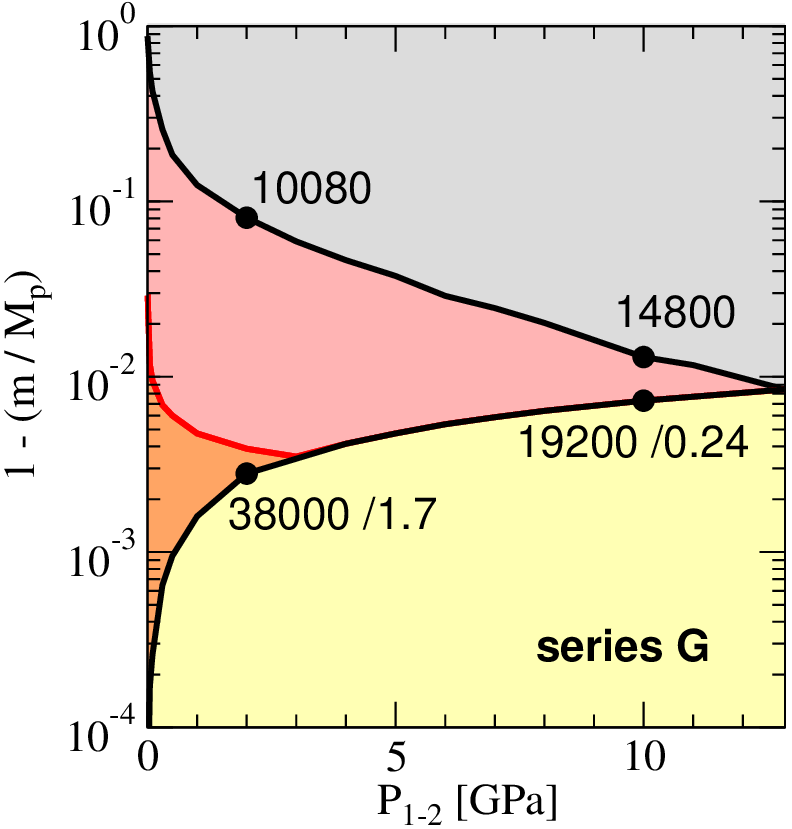}
\includegraphics[angle=0,width=4cm]{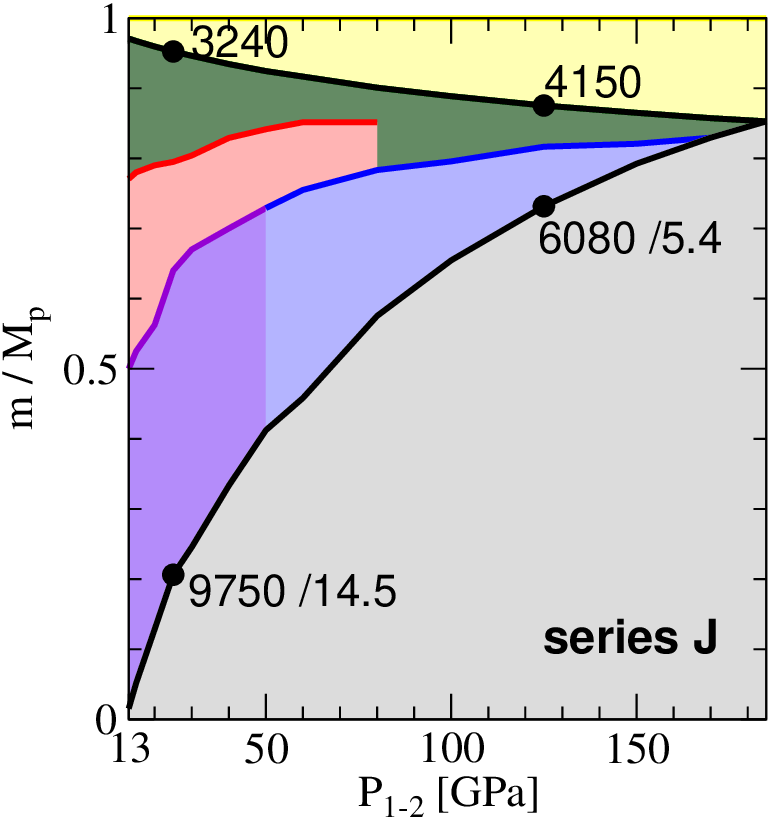}
\includegraphics[angle=0,width=4cm]{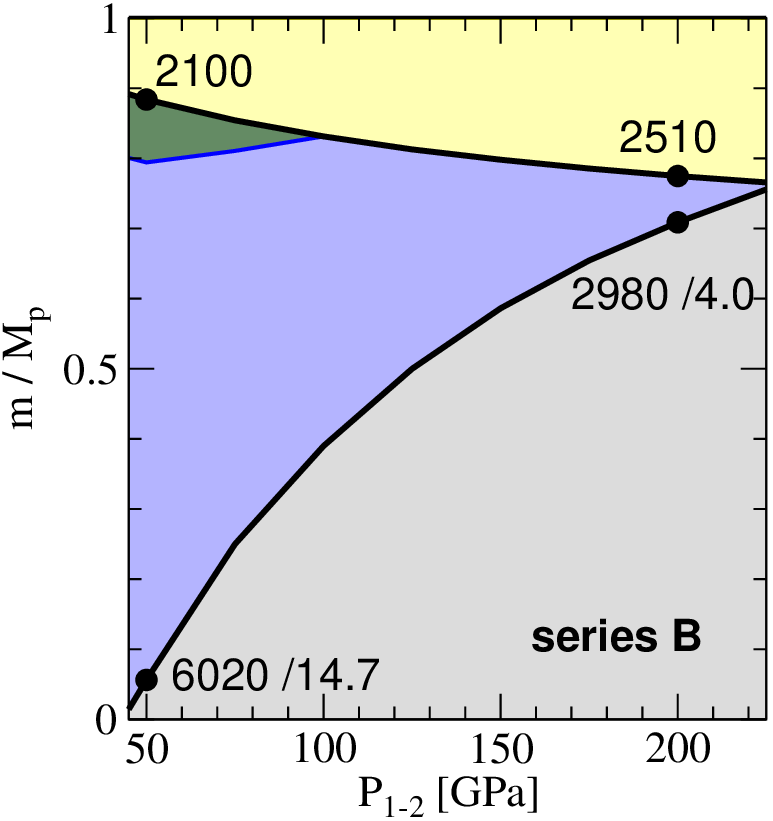}
\includegraphics[angle=0,width=4cm]{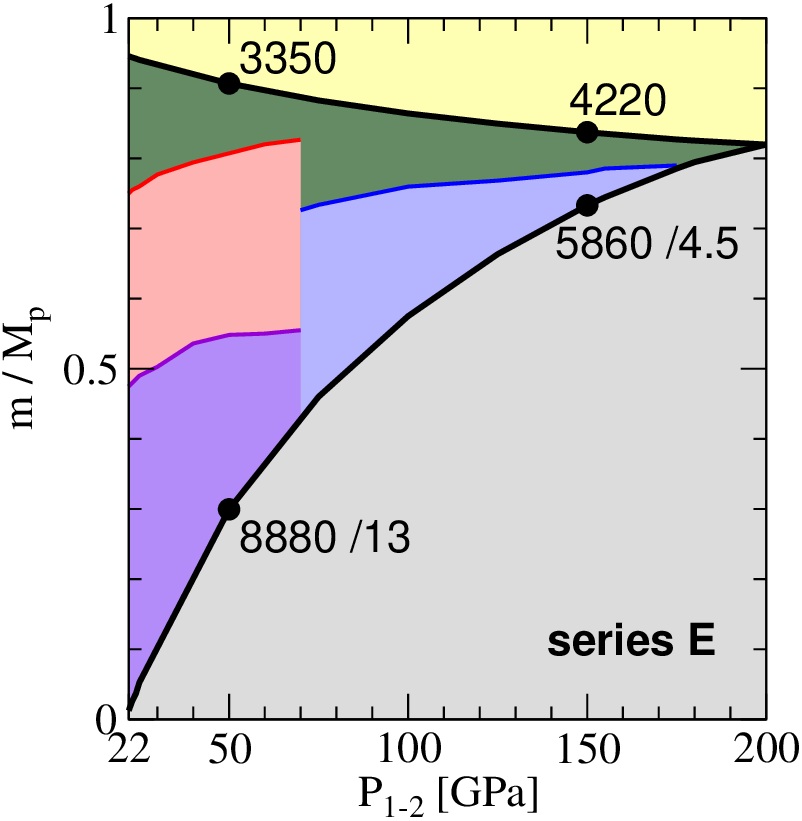}
\includegraphics[angle=0,width=4cm]{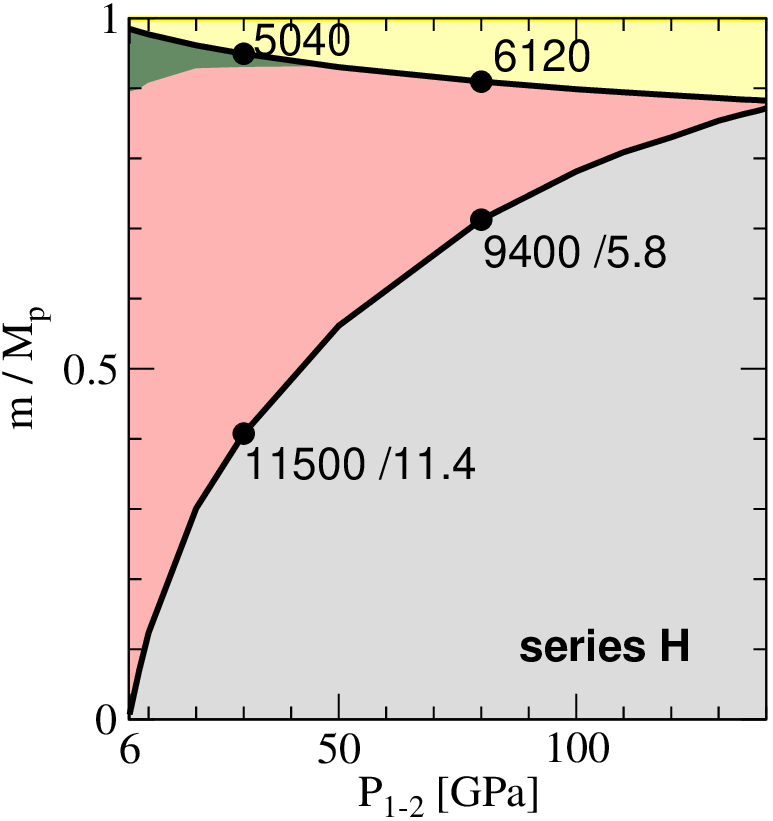}
\includegraphics[angle=0,width=4cm]{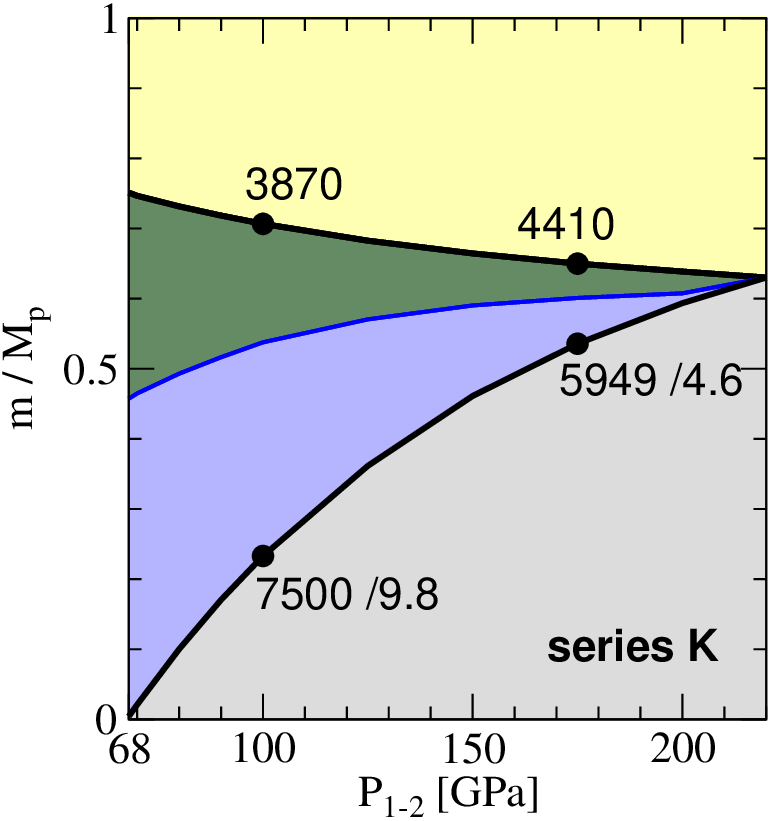}
\includegraphics[angle=0,width=4cm]{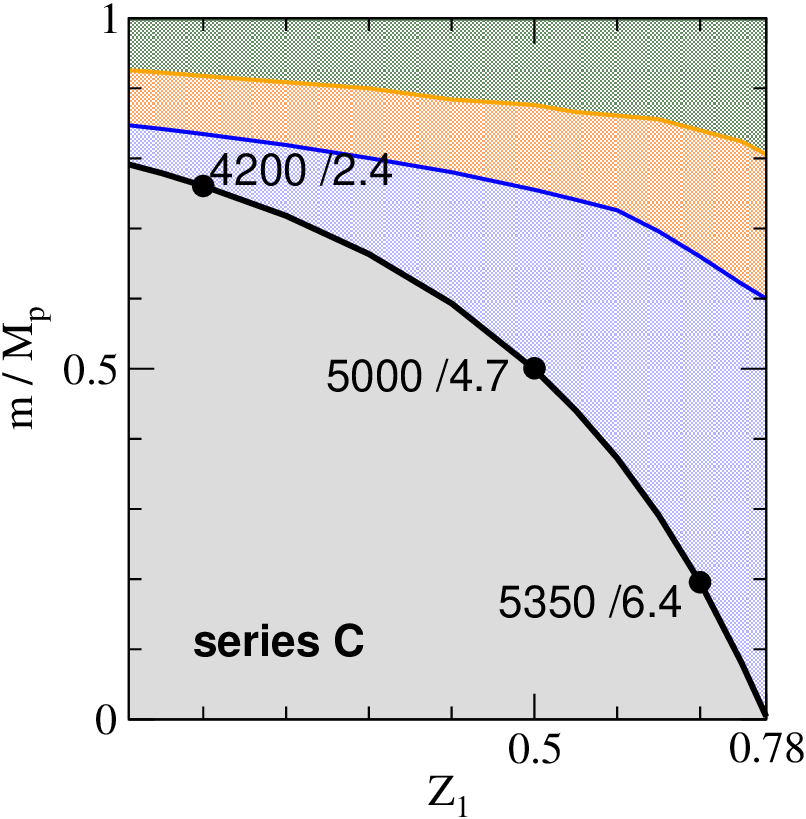}
\includegraphics[angle=0,width=4cm]{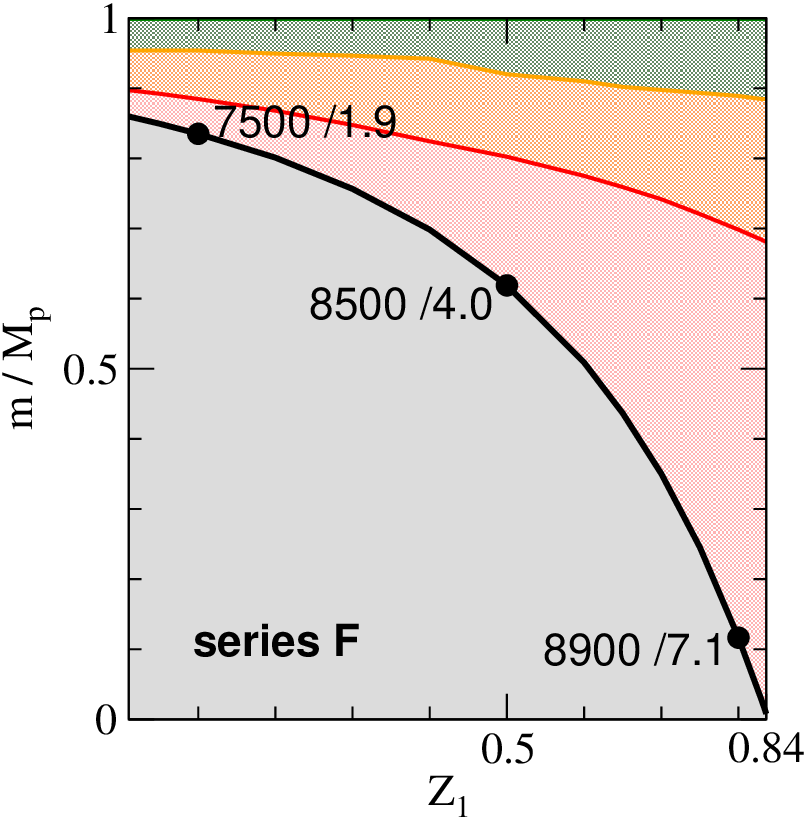}
\includegraphics[angle=0,width=4cm]{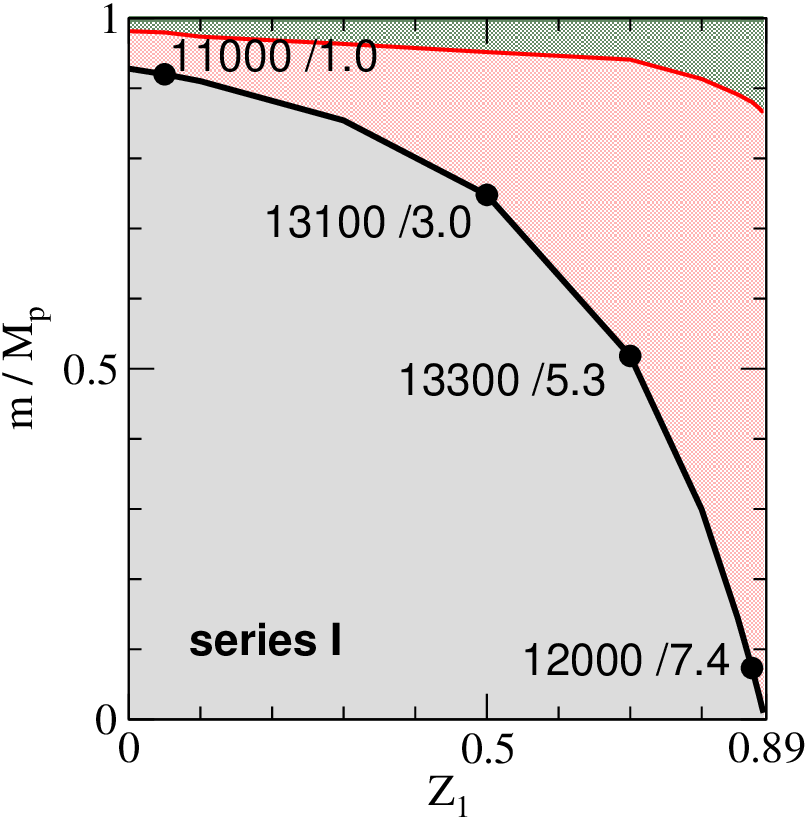}
\caption{\label{fig1}Mass distribution of the three components H/He (\emph{yellow}), water, 
and rocks (\emph{gray}) for the 11 series A-K described in Table~\ref{tab:series}. Each panel 
shows various models running vertically, which differ in transition pressure $P_{1-2}$ 
in the case of 3L models, and in envelope water mass fraction $Z_1$ for 2L models 
(series C, F, and I). \emph{Thick black lines} show layer boundaries, and numbers close to 
\emph{filled black circles} denote temperature in K and pressure in Mbar. Water is subdivided 
into five colors coding fluid molecular and fluid dissociated water (\emph{green}), plasma 
(\emph{red}), superionic water (\emph{blue}), an undissolved region in the water phase diagram 
close to the superionic phase (\emph{violet}), and fluid dissociated water (\emph{orange}) 
in series C, F, and G. In series C, F and I, the H/He mass fraction can be obtained via 
$(1-Z_1)\times(1-M_{core}/M_p)$. In series G, the center of the planet is at the top of the 
figure while for all other series, the center is at the bottom.}
\end{figure*}

We study the effect of surface temperature, atmospheric temperature profile, 
and mean density on the bulk composition in Fig.~\ref{fig1}. Each panel portrays 
the mass distribution of the three bulk components H/He, water and rocks. For the 
3L models, panels show the composition in dependence on the transition pressure 
$P_{1-2}$ between the H/He envelope and the water envelope. Layers are separated. 
The water layer is subdivided in up to five regions indicating, in accordance with the 
water phase diagram, the fluid molecular and fluid dissociated phase, the dissociated 
phase (only series C, F, G), the plasma phase, the superionic phase, and a thin region 
around the first-order phase transition between plasma and superionic water. 
Due to the finite resolution of the water phase diagram, these water layer regions are 
accurate within $0.5\ME$.  For the 2L models (series C, F, I) panels show the composition 
in dependence on envelope metallicity $Z_1$.

For all series, models without water are possible as well as models without rocky core, 
constituting the limiting cases of each series. 
Isothermal surface layers extendig down to 1 kbar reduce the core temperature by a factor 
of 2 to 3 compared to the fully adiabatic case (compare e.g.~series A, 
where $T_{core}=9900-15000$~K and B, where $T_{core}=2700-6100$~K). The same effect is 
observed from the assumed uncertainty in $T_1$ (compare e.g.~series A and G). Pressures 
at the core-mantle boundary range from 1 Mbar for large cores up to 15~Mbar for small cores, 
and the water density increases up to 6.5~g~cm$^{-3}$.

For the fiducial $M$-$R$ values, the coldest models obtained have deep internal temperatures 
as low as $\approx 3000$~K at 2-4 Mbar.  According to the FT-DFT-MD-based H$_2$O phase diagram, 
water then adopts the superionic phase. This phase differs from water-ice phases VII and X 
occurring at temperatures below 2500~K at these pressures, in that protons are not confined to 
lattice positions but diffuse through the bcc oxygen lattice making the medium electrically 
conductive \citep{Redmer+10}. Despite the low minimum surface temperature of 300~K considered, 
we find no model where water is in an ice phase. The melting temperature is even lower for 
alloys of water ice mixed with other species at any given pressure, supporting our conclusion 
that GJ$\:$436b does not contain ice. Indeed, models with fully adiabatic profiles and 
those with $T_1\gg 700$~K only have a 0-3$\ME$ thin mass shell where water is fluid before it 
transits smoothly to the plasma phase because of rising temperature. 
For $T_1=700\pm 100$~K down to 1 kbar (e.g.~series E) it is sensitive to the extension 
of the H/He envelope, if deep interior pressures become high enough for superionic or plasma 
water to be the preferred phase. Eye-striking differences between the 3L models with and without 
isothermal atmosphere indicate that the temperature at $\approx 1$~kbar is an important quantity 
demanding accurate modeling of the atmosphere in order to narrow down the set of solutions. 
Series A models have $T_{1000}=1950$~K, series D models have $T_{1000}=4000$~K, and series G 
models have $T_{1000}=5500$~K.
The value of $T_{1000}$ strongly influences the amount of H/He required to match $R_p$: 
$M_{\rm H/He}/M_p$ decreases from 11$-$25\% in series B ($T_{1000}=300$~K) to 2$-$11\% in 
series H ($T_{1000}=1300$~K), further to 0.7$-$10\% in series A to 0.1$-$4\% in series D, 
and finally to 0.01$-$1\% in series G. 

Since the scale height of the atmosphere decreases with mean molecular weight 
\citep{MillerRicci+09}, outer H/He envelopes occupy a large volume and thus require a 
relatively low mass to account for the observed radius of planets. However, H and He 
are compressible materials so that, if mixed into the deep envelope where pressure rises 
up to 15 Mbar, the scale height of an H/He-water layer becomes less than that of separated 
layers. Therefore, 2L models generally require a higher H/He mass fraction at same boundary 
conditions. In series C, the H/He mass fraction is 20.6$-$35.7\%, in series F 13.8$-$19.6\%, 
and in series I 5.6$-$14.5\%, thus more than in the comparable cases B, E, and H.
 
Increasing the mean density within the observational error bars of $M_p$ and $R_p$ has a neglegible 
effect on bulk composition and internal $P-T$ profile, compare series J with E. On the other hand, 
the minimum mean density $\bar{\rho}_p=0.45\rm~g~cm^{-3}$ increases the H/He mass fraction from 
5$-$18\% to 25$-$37\%, compare series K and E. Furthermore, less dense models have slightly lower 
temperatures and pressures at given mass shell so that, for the atmospheric temperature chosen, 
water in the deep interior will be superionic in all series K models.

The bulk composition of all models presented here is $M_{\rm H/He}/M_p$ = 10$^{-4}$ - 0.37, 
$M_{\rm H_2O}/M_p$ = 0 - 0.9999, and $M_{rocks}/M_p$ = 0 - 0.99. While the H/He mass fraction
in models with $T_{1000}\geq 4000$~K can be as little as $10^{-4}-10^{-3}$, the H/He layer in 
such cases extends over 0.07 to 0.17~$R_p$ (0.30 to 0.72$\RE$). We conclude that H/He is an 
indispensable component of this planet. This is unlike GJ~1214b \citep{Charb+09}, where the
measured radius can be explained by the assumption of water steam atmospheres \citep{RogSea10b}.

Each model series can be compared to composition predictions from the planet formation 
models by \citet{Figueira+09}, who investigate the composition of a GJ$\:$436b-sized 
planet with separated layers of rocks+iron, ices, and H/He similar to our 3L models. 
Their conditions are $(i)$~$M_{rocks}/M_p=0.45-0.7$, $(ii)$~$M_{\rm H/He}/M_p=0.1-0.2$, 
and $(iii)$~$M_{\rm H2O}/M_p=0.17-0.40$. Considering water representative of a mixture of 
elements with a mean density distribution similar to that of water, i.e.~an appropriate 
mixture of ices, rocks, and H/He, these conditions would relax to $(i')$~$M_{rocks}/M_p<0.7$, 
$(ii')$~$M_{\rm H/He}<0.2$, and $(iii')$~$M_{\rm H2O}/M_p >0.17$. We compare in the following
with the stricter conditions having in mind that more models might be possible if these
conditions are relaxed.
Apparently, no model of series A, D, and G satisfies $(i-iii)$, since high temperatures in 
the adiabatic outer envelopes act to decrease the mass density at given pressure level, 
and the pressure distribution with mass is predominantly a function of $M_p(R_p)$ alone. 
As a result, these models have thin H/He envelopes and large cores to account for $M_p(R_p)$.
In the coldest model series B, we see agreement with $(i-iii)$ if $P_{1-2}=115-160$~GPa. 
Such models have completely superionic water layers. In series E, we see agreement 
if $P_{1-2}=75-120$~GPa. Water forms an ionic fluid in a thin shell before it transits 
to the superionic phase. This is similar to the interior of Uranus and Neptune 
\citep{Redmer+10}, whereas the rocky core mass fraction of 0.45$-$0.6 in these models 
is much greater than in Uranus $(\leq 0.18)$ or Neptune $(\leq 0.24)$.

In series H, we see agreement only if $P_{1-2}\approx 80$~GPa. However, the water shell is 
in the plasma phase, so that strict separation from the H/He envelope might not be realized. 
We conclude that models of series H are unlikely. The same argument holds for series A, D, and G. 
With series C and F, strict and relaxed conditions give the same possible range of models. 
We see agreement, if $Z_1\approx 0.6$ (F: 0.4$-$0.5), while the H/He content becomes 
too large for higher envelope metallicities, and the total water mass fraction too small for 
lower metallicities. Models of series C are cold enough for water to be superionic. 
Whether a lattice structure of the oxygen subsystem or homogeneity throughout such an envelope 
can be maintained remains to be investigated. With series I, we see agreement if $Z_1=0.55-0.7$. 
These models have $T_{core}\approx 13,300$~K and $P_{core}=3-5.3$~Mbar, where water is in a 
plasma state, hence likely to be mixed with the H/He component. The same holds for the allowed 
series F models, where $T_{core}=8100-8500$~K. Thus these models can be considered self-consistently.
With series J, we see agreement if $P_{1-2}=65-110$~Mbar, where water close to the core is 
superionic justifying the assumption of a water shell. The fluid fraction of the water layer might 
be mixed into the H/He envelope. The same holds for series E ($\bar{\rho}=1.69\rm~g~cm^{-3}$), 
which differs from series J ($\bar{\rho}=2.22\rm~g~cm^{-3}$) only in mean mass density $\bar{\rho}$.
The similarity of the range of series E and J models implies that the upper limit of this 
observational quantity is sufficiently well known.

For series K models, the H/He content is clearly more than predicted by \citet{Figueira+09}, 
since they investigate the formation of a planet with $R_p\leq 4.4\RE$, whereas in series K, 
$R_p=5.35\RE$ ($\bar{\rho}=0.45\rm~g~cm^{-3}$), requiring a larger fraction of light elements. 
However, \citet{Lecavelier07} predicts complete evaporation of GJ$\:$436b within 5~Ga,
if $\bar{\rho}_p<0.74\rm~g~cm^{-3}$. While contrary positions exist \citep{Erkaev+07}, 
theoretical mass loss rates offer a way to tighten the uncertainty in age and maximal
radius of GJ$\:$436b.

According to the models by \citet{Spiegel+10}, the atmosphere is isothermal at $T=1300$~K 
between about 1 and 100$-$1000~bar, corresponding to our series H and I. From the discussion 
above, inconsistency with the water phase diagram of series H indicates that water is at 
least partially mixed within the H/He envelope. The extreme case of complete mixing between 
H/He and H$_2$O is realized in series I. Models of that series with $Z_1=0.55-0.70$ give 
the best overall agreement with available constraints. Those models have an H/He mass 
fraction of 0.13$-$0.145, while core mass and water mass fraction cover the range proposed 
by \citet{Figueira+09}. Alternative constraints may become available from improved formation 
models and atmosphere models in the future and be applied to our broad set of series so that 
improved estimates of the bulk composition can then be derived from Fig.~\ref{fig1}.

Our models comprise those by \citet{RogSea10a}, who find an H/He mass fraction of 2$-$14.5\%. 
Temperature-effects of our water equation of state tend to enhance the volume of a mass 
shell at given pressure level, requiring less H/He to be added to reproduce the large radius. 
At typical internal conditions $P\sim 2$~Mbar and $T\sim 6000$~K in GJ$\:$436b, the 
temperature effects enhance the volume by $\approx 20\%$ compared to water at 300~K 
\citep{French+09}. For this reason, we are able to find models with very low H/He mass 
fractions ($\approx 10^{-3}M_p$).

\subsection{Love number, core mass, and metallicity of GJ~436b}\label{ssec:res_k2}

\begin{figure}[hbt]
\centering
\resizebox{\hsize}{!}{\includegraphics[angle=0,width=0.5\textwidth]{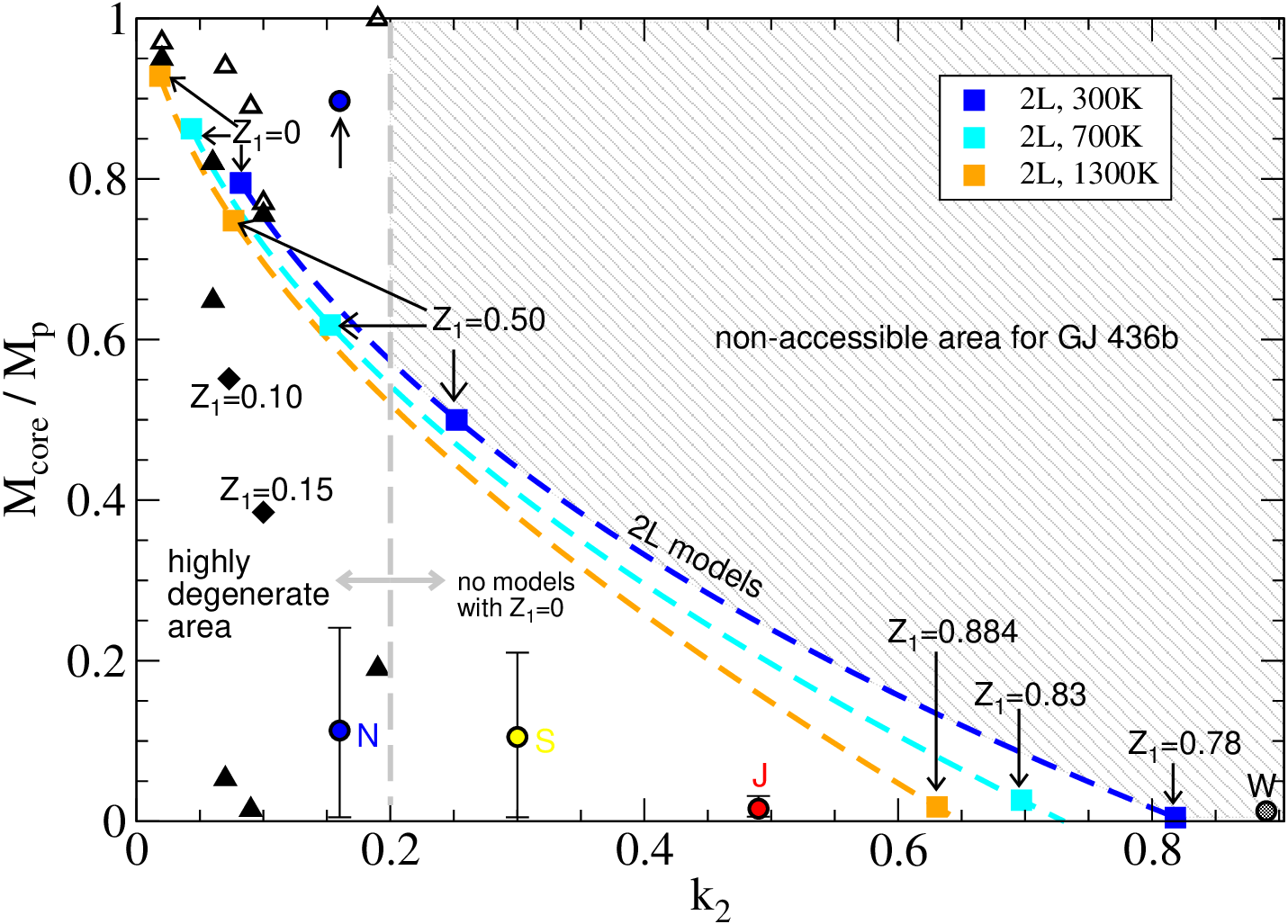}}
\caption{Love number $k_2$ and core mass of two-layer models (\emph{squares} connected by 
\emph{dashed lines}), three-layer models with metal-free envelopes (\emph{triangles}), and
three-layer models with low-metallicity envelopes (\emph{diamonds}) of GJ~436b. 
Color-coded circles are for interior models of Jupiter (\emph{red}), Saturn (\emph{yellow}), 
and Neptune (\emph{blue}), and a $20\ME$ water planet (\emph{black}).
Adding the mass of the water layer to the mass of the rocky core shifts 3L models solutions
from \emph{filled} to \emph{open triangles}, and the Neptune model to the position as indicated
by an arrow. For each of the \emph{dashed lines} (which are for different isothermal atmospheres 
of 300~K (\emph{blue}), 700~K (\emph{cyan}), and 1300~K (\emph{orange})), the squares indicate 
solutions for $Z_1=0$, 0.5, and the highest possible value when $M_{core}=0$. 
On the left hand side of the \emph{gray dashed line} at $k_2=0.2$, the models are highly
degenerate, and on the right hand side, no metal-free envelope models were found.}
\label{fig:k2}
\end{figure}

For various models we calculated the Love number $k_2$. This parameter is known to 
measure the \emph{central condensation} or, equivalently, homogeneity of matter in a 
nearly spherical object \citep{ZT78} with an upper limit of 3/2 for a sphere of constant 
density and a lower limit of zero for Roche-type models. Jupiter's theoretical value is 
$k_{2,J}\sim 0.49$ \citep{RW09} indicating a close to $n=1$ polytropic density profile, 
which is confirmed by our calculations. 
For Neptune, we find $k_{2,N}=0.16$ indicating stronger central condensation, for Saturn 
we find intermediate values around $k_{2,S}=0.31$, and for an artificial $M_p=20\ME$ water 
planet we find $k_{2,W}=0.89$, in agreement with water being less compressible than 
the H/He dominated envelopes of the solar giant planets. Finally for GJ$\:$436b, 
we find $k_2$ ranging from 0.02 to 0.82. These results are displayed in Fig.~\ref{fig:k2}, 
together with the uncertainty in core mass of these planets.

Models of GJ$\:436$b with $k_2=0.02-0.2$ are accompanied by any core mass between 0 and 
$\approx 0.9M_p$: a measurement of $k_2$ in this regime would not help to constrain 
the interior any further. All our calculated 3L models with metal-free or low-metallicity 
H/He envelopes fall into this degenerate regime. Beyond the region of strong degeneracy 
at $k_2>0.2$, the uncertainty in $M_{core}$ for an observationally given $k_2$ 
shrinks with $k_2$, so that upper limits for the uncertainty in core mass can be derived, 
if $k_2$ and the atmospheric temperature profile are known. Those models are necessarily 
two-layer models with one homogeneous envelope. Assuming for instance an atmospheric 
temperature of 1300~K, then any redistribution of metals from the outer part of the 
envelope to the inner part of the envelope would (if the central condensation is to be 
kept at constant $k_2$ value) require a compensating transport of matter from the core upward, 
making the core mass decrease and drop below the upper limit for that atmospheric
temperature at any given $k_2$. This upper limit in core mass increases with decreasing 
atmospheric temperature, since dense envelopes act to reduce the density gap between envelope 
and core, which would reduce the central cendensation ($k_2$), if it was not compensated for 
by a more massive core. 
If we adopt 300~K as coldest possible temperature of an isothermal atmosphere, then the area 
above the blue dashed curve in Fig.\ref{fig:k2} is not accessible by present GJ$\:$436b. 
The high-core mass end of that curve corresponds to a zero-metallicity envelope. At the 
high-core mass end, however, the solutions lie in the area of large degeneracy, where besides 
the mass the mean density of the core strongly affects $k_2$, and the behavior of solutions 
is no longer a simple function of envelope metallicity and temperature.

At Saturn's theoretical $k_{2,S}=0.31$, the uncertainty in core mass is $\approx 0.4M_p$, 
it is $\approx 0.2M_p$ at Jupiter's $k_{2,J}=0.49$, and it is $\leq 0.15M_p$ for $k_2>0.6$. 
Such high $k_2$ values, if measured, imply high outer envelope metallicities for this planet. 
For 2L models, we find $0.55<Z_1<0.78$ in $300~K$ cold atmospheres, 
and $0.7<Z_1<0.884$ in $1300~K$ hot atmospheres, where $k_2$ is a sensitive function 
of $Z_1$. An observational $k_2$ value between 0.3 and 0.8 would help constrain envelope 
metallicity and core mass.

Furthermore, a superionic water layer can be regarded in the traditional manner as a 
former ice layer of the initial core that warmed up during formation through mass accretion 
and contraction. Adding the mass of the water layer to the underlying rocky core results 
in significantly higher core masses. The case of Neptune illustrates the general difficulty 
of applying the label \emph{core mass} to metal-rich planets with a layered structure.

\subsection{$M$-$R$ relations for water planets}\label{ssec:MRrels}

\begin{figure}[hbt]
\centering
\resizebox{\hsize}{!}{\includegraphics[angle=0,width=\textwidth]{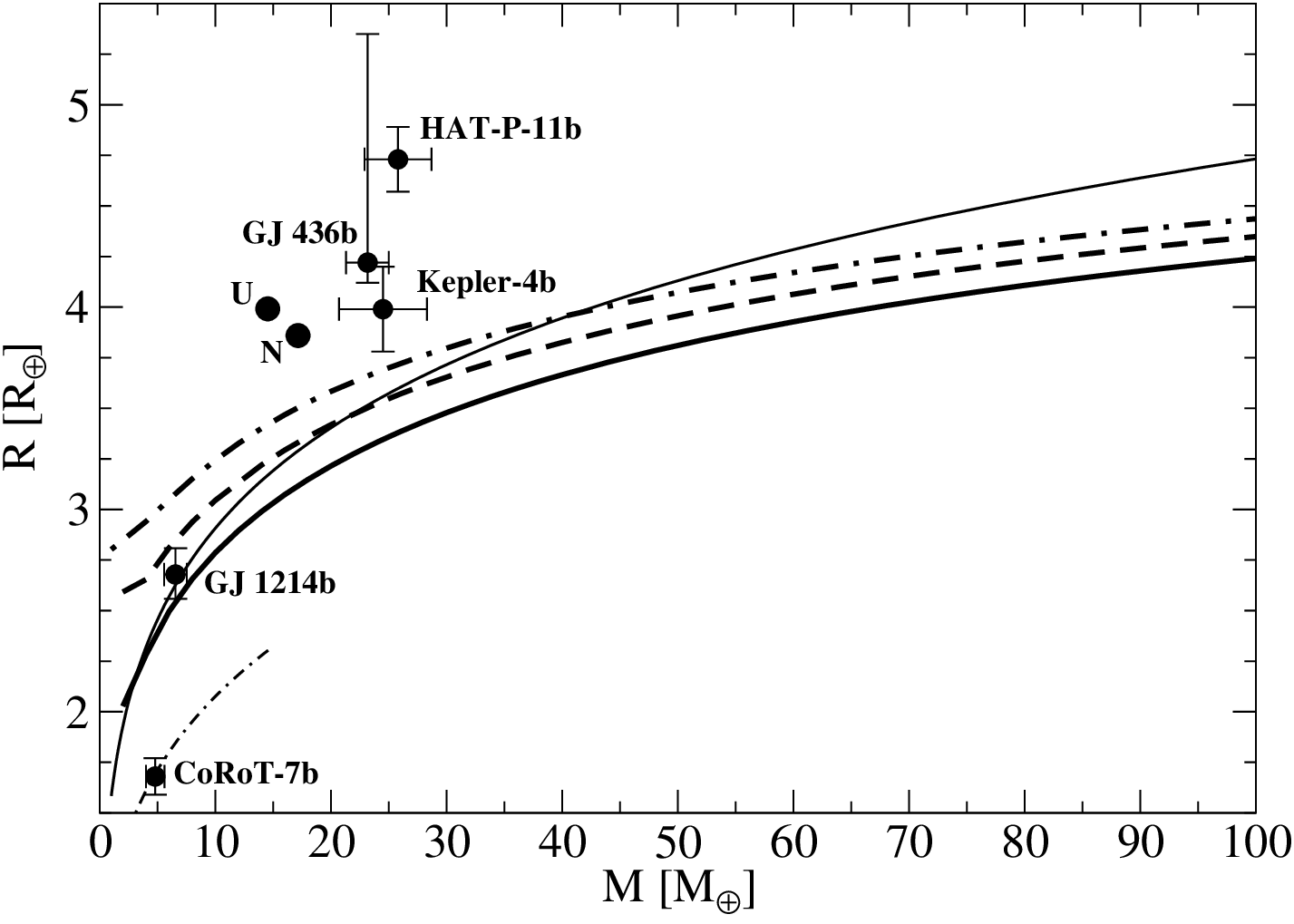}}
\caption{Mass-radius relationships of water planets in comparison with Uranus (U), Neptune (N), 
the three known Neptune-mass planets, and the two known super-Earth mass planets GJ$\:$1214b
and CoRoT-7b. 
\emph{Solid line}: $T_1$=1000~K, isothermal atmosphere, $M_{core}=$0.25$M_{\oplus}$; 
\emph{dashed line}: same as solid line but $T_1$=2000~K; \emph{dot-dashed line}: 
same as solid line but adiabatic atmosphere; \emph{thin solid line}: fit-formula for 
water planets from \citet{Fortney+07b}. The \emph{thin dashed-dot} line is the theoretical $M$-$R$
relation for Mg-silicate planets from \citet{Valencia+10}.}
\label{fig:mr}
\end{figure}

Figure~\ref{fig:mr} shows $M$-$R$ relations of water-planets with a small ($0.25\ME$) 
rocky core, using H$_2$O-REOS. We present the $M$-$R$ relation for profiles with an  
isothermal atmosphere down to 1 kbar of 1000~K, 2000~K, for fully adiabatic profiles 
starting with $T_1$=1000~K, and, for comparison, for the fit-formula from \citet{Fortney+07b} 
for pure adiabatic water planets based on a finite temperature correction to a zero-K 
isotherm of water. Our $M$-$R$ relations for $T_1=1000$~K  can be fitted with the formula  
\begin{equation}\label{eq_MRfit}
	R(M)=a_0 + a_1*\log M + a_2*\log M^2 + a_3*\log M^3
\end{equation}
with $a_0$=1.6586, $a_1$=0.9950, $a_2$=0.1549, $a_3$=0 for the solid line and $a_0$=2.8210, 
$a_1$=$-0.2928$,\, $a_2$=0.9037, $a_3$=$-0.1760$ for the dashed-dot line in Fig.~\ref{fig:mr}. 
This uncertainty in the atmospheric temperature profile (isothermal or adiabatic) obviously 
influences the radius by~$\sim 0.5\RE$. 

In the 5$-$50~M$_{\oplus}$ planet mass region, previous and current $M$-$R$ relations 
are similar with a flatter radius dependence in the latter case indicating a smaller 
polytropic exponent $\gamma$. In particular, the interior profiles of our water planets 
can be approximated well by $\gamma=2.8$ for $2<\rho<4\rm~g~cm^{-3}$ and $\gamma=2.4$ for 
$\rho>4\rm~g~cm^{-3}$.
The resulting theoretical radii of water planets are below the observed radius of the three 
currently known Neptune-mass planets GJ$\:$436b, HAT-P-11b \citep{Bakos+10}, and Kepler-4b 
\citep{Borucki+10}, indicating the presence of low-Z material in these planets. 
In case of warm atmospheric temperatures of 1000~K or more, an outer H/He envelope 
that contributes in radius $\sim 15$\% may contribute $\ll 1$\% in mass. The only known planet 
whose observational parameters $M_p$, $R_p$, and $T_{eq}$ can be explained by the assumption 
of a water-dominated water+rock composion without H/He \citep{RogSea10b} is GJ$\:$1214b, 
whereas for the super-Earth CoRoT-7b \citep{Leger+09}, $M_p=4.8\pm 0.8$, a water mass fraction 
$<10$\% and an H/He mass fraction $<0.01$\% are predicted by evolution calculations 
that include the energy-limited mass loss caused by stellar irradiation \citep{Valencia+10}.

\section{Discussion}\label{sec:discussion}
 
\subsection{A secondary planet} 

To explain the high eccentricity $e=0.15$ of GJ$\:$436b, the presence of a 
secondary planet has been suggested \citep{Ribas+08} but not yet confirmed 
by measurements \citep{Bean+08}. While observations exclude a resonant perturber 
in 2:1 resonance down to lunar mass, the presence of a secular perturber at distant 
orbit in apsidal alingment with GJ$\:$436b is still an appealing possibility 
\citep{Batygin+09a}. If detected, the scheme presented in \citet{Batygin+09b} allows 
determination of $k_2$, if the planets are on coplanar orbits \citep{Mardling10}. 
We predict $0.02\leq k_2 <0.82$.

If future measurements deny there is a secular perturber, a (minimum) Love
number can be estimated by requiring sufficiently weak tidal interactions
between planet and star that would allow for the planet's presence at observed orbital 
distance and eccentricity.
Using the common low-eccentricity approximation of the tidal evolution equations, 
\citet{Jackson+08} find that an effective constant tidal $Q'$ value of $10^{6.5}$ 
results in an initial $a_0-e_0$ distribution that best matches the observed distribution
of tidally unevolved planets.
With $Q'=3Q_p/2k_2$ (in the approximation of incompressible homogeneous matter; 
see \citealp{GoldSot66} and \citealp[][\S123]{Love1911}) 
where $Q_p^{-1}$ is the tidal dissipation function, and with $Q_p=Q_N=(0.9-33)\times 10^4$ 
\citep{ZhangHam08,BanMur92} as proposed for Neptune, this implies $k_2=0.0043-0.156$. 
We find $k_2\geq 0.02$. 
This raises the possibility that the $Q$ value of GJ$\:$436b is close to Neptune's upper 
limit, making the eccentricity a consequence of weak tidal interaction with the star. 

However, according to \citet{Jackson+08} the initial eccentricity of GJ$\:$436b would 
have been $e_0 \approx 0.6$, which has been demonstrated to lie beyond the range of 
applicability of the low-eccentricity approximation, as the theoretical tidal evolution 
timescale can decrease by several orders of magnitude when the exact orbital evolution
theory is used \citep{Leconte+10}. A re-calculation of GJ$\:$436b's tidal evolution 
would therefore be helpful in order to obtain present-time $(Q',k_2)$ pairs that reproduce 
the observed orbital parameters, thereby clarifying the need for an external perturbing body.

\subsection{Constraints from formation and atmosphere profiles}

Among the broad sets of models presented here, those that are consistent with available 
constraints from formation theory, model atmospheres, and solubility of fluid water in 
H/He have a warm outer envelope with water mass fraction 55$-$70\%, and a rocky core. 
At such large envelope heavy element abundances, $k_2$ begins to be sensitive 
to core mass. On the other hand, measuring $k_2>0.3$ would either imply a violation of 
the condition from formation theory $M_{rock}/M_p>0.45$ or indicate the presence of rocks 
in the envelopes (partially replacing water). 
In the outer part of the envelope, water is in a supercritical fluid state and then transits 
to the plasma phase because of rising temperatures. This supports the assumption that water 
is mixed into the H/He envelope in the two layer models instead of forming a separate 
layer as in the three layer models. Depending on the process of formation, however, 
a separation into few distinct layers with different water fractions can be possible.
Colder models where water is in a superionic state at 3000-6000 K are not consistent with the 
$\approx 1300$~K hot model atmospheres by \citet{Spiegel+10} or the recent model
atmospheres by \citet{Lewis+10} which predict a radiative layer between 1 and 100 bar
at 1100 to 1200~K depending on metallicity. High-pressure ice phases occur
at even lower temperatures, hence in none of our models. Considering water as a proxy 
for metals, higher H/He mass fractions are possible than the 0.13-0.145 of the preferred  
models. The broad set of models presented here demonstrates the importance of accurate model 
atmospheres. Neglecting the available constraints from formation and atmosphere models, the 
set of solutions satisfying only $M_p$, $R_p$ and consistency with $\Tpeff$ includes 
models with $M_{\rm H/He}/M_p=10^{-3}$ and $M_{\rm H_2O}/M_p=0.999$.

\subsection{Constraints from atmosphere observations}

For GJ$\:$436b, an observational value $k_2>0.5$ implies a rocky core mass below $0.2M_p$ 
and an envelope metallicity over 50\%. An observational value $k_2>0.2$ would 
imply $M_{core}<0.55M_p$, and $Z_1$ above $\approx$ 40\% if the envelope is homogeneous.
While homogeneous envelope models are  supported by the hot internal temperatures 
and miscibility of water and hydrogen, we cannot be sure about the absence 
of envelope layer boundaries, so can only obtain upper limits for the core
mass for any given $k_2$ value. In order to break this degeneracy further, an estimate
of the outer envelope metallicity is required.

For solar system giant planets, atmospheric heavy element abundances are only known 
within a factor of a few for some species that do not condense at optically thick deeper 
levels. The abundances are believed to be representative of the outer envelope because 
of vertical mixing still at high altitudes below 1 bar. 
Irradiated extrasolar planets, such as GJ 436b however, become optically thick at pressures 
several orders of magnitude lower than the onset of the convective envelope.
Even in the fortunate case of measured atmospheric abundances, representative metallicity 
determinations might require including non-equilibrium chemistry, cloud formation, 
vertical mixing beyond mixing-length theory, and other effects.
On the other hand, current theoretical and observational progress is encouraging. 
Transmission and emission spectra predicted by state-of-the-art model atmospheres 
exhibit various observational signatures in dependence on metallicity.
\citet{Chabrier+07} found a 30\% decrease of a hot Jupiter's emission at $4.5\mu$m due 
to encanced absorption by CO when the overall metallicity was enhanced from 1x to 5x solar.
For super-Earth atmospheres, \citet{MillerRicci+09} found large hydrogen fractions (low
mean molecular weight) to map on a detectable variability of the transit radius with 
wavelength and a rapid decrease in variability with mean molecular weight.
For GJ$\:$436b in particular, \citet{Lewis+10} developed a three-dimensional atmospheric
circulation  model and predict metal-enhanced atmospheres to reveal themselves in an 
orbital phase dependency of spectra potentially detectable during future missions.
In addition, secondary eclipse measurements \citep{Stevenson+10} already provide
hints of a strongly (30-50x solar) metal-enhanced atmosphere \citep{Lewis+10}.

\subsection{Comparison with Uranus and Neptune}

Interior models of Uranus ($M_p=14.5\ME$) predict that at least 21\% by mass is H/He, 
and at least 15\% in Neptune ($M_p=17.1\ME$) \citep{ForNet09}.
This is consistent with the preferred model values for GJ~436b (13$-$14.5\%) considering that
some fraction of water as assumed in the model calculations could indeed be rocks and H/He 
in the envelope of either of the planets. On the other hand, since only lower limits of the 
H/He to ice and rock ratios in these planets can be derived, their composition can also 
be very different from each other.
Irradiation by the close-by parent star may rise the core temperatures of preferred 
GJ$\:$436b models significantly above those of Uranus and Neptune, where adiabatic models 
predict $T_{core}=6000-7000$~K, hence water in the superionic phase. Methane has been 
suggested as separating into metallic hydrogen and diamond \citep[e.g.][]{Hirai+09} at those 
deep interior conditions, so that the interior of those planets might be organized very different 
by that from GJ$\:$436b, even in the case of similar bulk composition. Interior models of Uranus 
and Neptune require supersolar heavy element abundance in the outer envelope and some admixture 
of H/He in the deep interior, which can be represented by at least two envelopes with different 
compositions. Such a structure is bracketed by our 2L and 3L models. The observed low heat flux 
of Uranus and, to a lesser extend, of Neptune gives further evidence of layer boudaries, 
or more general, inhomogeneities in these planets \citep{Hubb+95}. Unfortunately, 
the age of GJ$\:$436 is poorly known, but investigations of the cooling behavior of various 
interior models can help further rule out or strenghten some of the models presented here.

\subsection{Conclusions}

Applying H-, He-, and H$_2$O-REOS for hydrogen, helium, and water to generate three-layer models 
(H/He envelope, water envelope, rocky core) and two-layer models (H/He/H$_2$O envelope, rocky core) 
of GJ~436b, we find interior models ranging from large rocky cores with significant H/He layers 
to almost pure water worlds with thin, but extended H/He atmospheres. If water occurs and the 1~kbar 
temperature is not much higher than 700~K, this exoplanet constitutes the third natural 
potential object for realyzing superionic water, besides the deep interior of Uranus and  
Neptune. In the H/He envelope, the temperature rises rapidly such that we do not find high-pressure 
ice phases in the inner envelope. Indeed, if we assume temperature profiles in good agreement with 
model atmospheres, water in the interior of GJ$\:$436b transits from fluid to plasma, favoring an 
outer envolope with strong water enrichment instead of a separation into two distinct envelopes.

The Love number $k_2$ of different possible compositions with metal-free outer H/He envelopes 
varies from 0.02 to 0.2. In this $k_2$ regime, the core mass depends non-uniquely on $k_2$
and cannot be derived from a measurement of $k_2$. An observationally derived value $k_2>0.2$ 
(0.3; 0.5) would imply $M_{core}<0.55M_p$ ($M_{core}<0.4M_p$; $M_{core}<0.2M_p$) 
and a high outer envelope metallicity.

The radius of GJ$\:$436b is larger than predicted for warm-water planets. Even in case of the 
warmest models found, a small H/He fraction of $10^{-3}M_p$ is necessary to account for the radius.
Calculating the cooling behavior on the basis of a proper treatment of the atmosphere might help  
to better constrain the interior of this planet. 

With this work we want to motivate both accurate 
transit-timing observations of this planet that aim to clarify the existence of a secondary planet
and that may allow derivation of $k_2$ and transmission spectra measurements aimed at determining 
the atmospheric metallicity.

\begin{acknowledgements}

We acknowledge discussions with P.H. Hauschildt, M. French, and J.J. Fortney and 
are grateful to D.J. Stevenson for his introduction to Love numbers and mentorship 
to one of us (UK). We thank the anonymous referee for his comments which extraordinarily 
helped to reshape and improve the paper.
This work was supported by the DFG within the SFB 652 and the grant RE 882/11-1 
and by the HLRN within the grants mvp00001 and mvp00006. We thank the Rechenzentrum 
of the U Rostock for assistance.
RN acknowledges general support from the German National Science Foundation 
(Deutsche Forschungsgemeinschaft, DFG) in grants NE 515/13-1, 13-2, and 23-1, 
as well as support from the EU in the FP6 MC ToK project MTKD-CT-2006-042514.
\end{acknowledgements}

\bibliography{ms11985bib}
\end{document}